\documentclass[11pt,a4paper]{article}

\usepackage{amssymb}
\usepackage{amsmath}
\usepackage{amsfonts}
\usepackage{graphicx}
\usepackage{a4wide}
\usepackage{microtype}
\usepackage{bbm}
\usepackage{slashed}
\usepackage{amsthm}
\usepackage{mathrsfs}
\usepackage{hyperref}
\usepackage{color}

\usepackage{pstricks}

\definecolor{darkred}{rgb}{0.8,0.1,0.1}
\hypersetup{
     colorlinks=true,         
     linkcolor=darkred,
     citecolor=blue,
}

\usepackage{comment}
\usepackage[margin=2.7cm]{geometry}

\usepackage{marvosym}

\theoremstyle{remark}

\def\un{1\kern-3pt \rm I}
\def\sl#1{\rlap{\hbox{$\mskip 1 mu /$}}#1}

\numberwithin{equation}{section}

\def\D{\mathcal{D}}

\def\1{\mathbbm{1}}
\def\g{\mathfrak{g}}

\def\b{\mathsf{b}}

\def\ssl#1{\rlap{\hbox{$ {\scriptstyle /}$}}#1}

\def\a{\alpha}
\def\b{\beta}
\def\d{\delta}
\def\e{\epsilon}
\def\ve{\varepsilon}
\def\f{\phi}
\def\vf{\varphi}
\def\g{\gamma}

\def\j{\Psi}

\def\l{\lambda}
\def\m{\mu}
\def\n{\nu}

\def\r{\rho}
\def\s{\sigma}

\def\x{\xi}

\def\D{\Delta}
\def\F{\Phi}
\def\G{\Gamma}

\def\O{\Omega}

\def\S{\Sigma}

\def\cf{{\cal F}}

\def\co{{\cal O}}

\def\cs{{\cal S}}

\newcommand{\ov}{\overline}

\newcommand{\wh}{\widehat}

\newcommand{\pa}{\partial}

\def\sl#1{\rlap{\hbox{$\mskip 1 mu /$}}#1}

\def\I{\leavevmode\hbox{\small1\kern-3.8pt\normalsize1}}

\newcommand{\ie}{{\it i.e.,\ }}
\newcommand{\eg}{{\it e.g.,\ }}

\title{%
\bf{Symanzik-Becchi-Rouet-Stora lessons on renormalizable models with
broken symmetry: \\ the case of Lorentz violation \\
{\normalsize(in memoriam of Raymond Stora -- 1930-2015)}}
}

\author{%
 Oswaldo M. Del Cima, Daniel H.T. Franco and Olivier Piguet \vspace{4mm}\\
      Universidade Federal de Vi\c cosa, Departamento de F\'\i sica\\
      Av. Peter Henry Rolfs s/n, Campus Universit\'ario,\\
      Vi\c cosa, MG, Brasil, 
      CEP:36570-900.\vspace{4mm}\\
{\small emails: \texttt{oswaldo.delcima@ufv.br}~,~\texttt{daniel.franco@ufv.br}~,~\texttt{opiguet@pq.cnpq.br}}
}

\date{\today}

\begin{document}

\maketitle

\begin{abstract}
In this paper, we revisit the issue intensively studied in recent years on the 
generation of terms by radiative corrections in models with broken Lorentz symmetry.
The algebraic perturbative method of handling the problem of renormalization of the theories 
with Lorentz symmetry breaking, is used. We hope to make clear the Symanzik's aphorism:
``{\it Whether you like it or not, you have to include in the lagrangian all counter terms consistent with 
locality and power-counting, unless otherwise constrained by Ward identities.}''\footnote{The phrase 
was borrowed from text ``{\it Pedagogical Experiments in Renormalized Perturbation Theory,}'' by Raymond Stora.
Contribution to the Hesselberg Meeting on the Theory of Renormalization and Regularization, 
24 February -- 1 March 2002.} 
\end{abstract}

\section{\bf Introduction}
\label{Sec1}
\hspace*{\parindent}
The study of perturbative field models with symmetry breaking were investigated from the point of view of 
the theory of renormalization in the pioneering work of Symanzik~\cite{Sym1,Sym2} and treated in a way that we can 
consider as definitive, by Becchi-Rouet-Stora~\cite{Stora1,Stora2,Stora3}. However, several recent works, dealing 
in particular with field theories with Lorentz symmetry breaking, do not consider very carefully how the symmetry is broken, 
not taking into account the requirements that  Symanzik-Becchi-Rouet-Stora have shown to be necessary. In this 
article we intend to write an updated review of the problem of renormalization of the theories with Lorentz symmetry 
breaking. All our analysis will be based on a general iterative scheme called 
``{\it Algebraic Renormalization}''~\cite{piguet-rouet,PigSor,Man,PigDan}.\footnote{It should be emphasized that, 
based on the method suggested by the Epstein-Glaser construction, the algebraic method 
of renormalization was ``seeded'' in the Lecture Notes by Professor Raymond Stora, ``{\em Lagrangian Field Theory,}'' in 
Particle Physics, Proccedings of the Le Houches Summer School, 1971, and edited by C. De Witt and C. Itzykson, 
Gordon \& Breach, 1973.} In the algebraic approach, in order to study the renormalizability of models characterized by a 
system of Ward identities, without referring to any special regularization procedure, two steps must be followed. 
In the first step, for a power-counting renormalizable model, at the level of the radiative corrections, one investigates 
the preservation of the symmetries, or the determination of all possible anomalies. This amounts to find the solution of 
the cohomology of its symmetry group: trivial elements (co-boundaries) correspond to breakings which can be compensated 
by non-invariant counterterms, whereas the non-trivial elements are the possible anomalies. These cohomology conditions
are a generalization of  the Wess-Zumino consistency condition~\cite{Wess-Zumino} used in
order to compute the possible anomalies of the Ward identities in Yang-Mill thories. In a second step, we  check the stability 
of the classical action -- which ensures that the quantum corrections do not produce counterterms corresponding to the 
renormalization of parameters not already present in the classical theory.

Let us emphasize that the algebraic renormalization scheme is based on a set of theorems of renormalization theory,
collected under the name of ``{\it Quantum Action Principle}'' (QAP)~\cite{Lam,Lowenstein,Brenneke-Dutsch}. These theorems 
deal with the whole of Feynman graphs' combinatorics and integrability, so that explicit graph considerations are 
unnecessary -- unless one looks for explicit quantitative results for applications to physics, of course. Said Raymond:
``{\it Use the theorems!}\,''~\cite{Raymond}.

This article is divided as follows: in Section \ref{Sec2}, we give a short review on the renormalization of theories with explicit 
symmetry breaking. In Section \ref{Sec3}, as a toy model, we analyze the Lorentz symmetry breaking in a scalar field model. 
Then, in Section \ref{Sec4}, we revisit the quantum electrodynamics (QED) with violation of Lorentz and CPT symmetries. 
Among several issues, in particular using the BRST formalism, we reassess the possible generation of a Chern-Simons-like 
term induced by radiative corrections arising from a CPT and Lorentz violating term in the fermionic sector, a recurrent theme 
in the literature. It is important to emphasize that, concerning extended QED with a term which violates the Lorentz and CPT 
symmetries, most of the papers were devoted to discuss the gauge invariance of the model {\it only}, putting aside a more 
specific way how Lorentz invariance is broken. In Section \ref{Sec5}, we finish with some reflections on the important works of 
Symanzik-Becchi-Rouet-Stora on renormalizable models with broken symmetry. This article is dedicated to Raymond Stora 
memory. The passion of Professor Stora for the fundamentals of Quantum Field Theory was what led him to become one of 
the leading researchers in the world of renormalization, culminating with the awards he has received: the Max Planck Medal 
(1998) and the Dannie Heineman Prize for Mathematical Physics (2009). The latter was a recognition of the important work 
he did with Carlo Becchi and Alain Rouet on a rigorous mathematical procedure for quantizing non-abelian gauge field theories, 
which is now known as BRST quantization.

\section{\bf Explicit symmetry breaking in a nutshell}
\label{Sec2}
\hspace*{\parindent}
In this section we present a sketch of the renormalization of models with broken symmetries, 
adapted from the works of Symanzik~\cite{Sym1,Sym2} and Stora and collaborators~\cite{Stora1,Stora2,Stora3}.
As in these works, we restrict ourselves to theories which are power-counting renormalizable and symmetries 
which are realized linearly.   

First, suppose that a set of field transformation laws is given, infinitesimally, by
\[
i \bigl[Q_\alpha,\Phi\bigr]=\delta_\alpha \Phi\,\,,
\]
where $\Phi$ is a field or multiplet of fields transforming in a
specific way under a symmetry group $G$. The charges $Q_\alpha$ form 
a basis of the Lie algebra of $G$, satisfying commutation rules
\[
\bigl[Q_\alpha,Q_\beta\bigr]=i f_{\alpha \beta \gamma} Q_\gamma\,\,.
\]
We may translate the transformations into the language of functional differential operators
\[
W_\alpha := -i \int d^Dx\,\,\delta_\alpha \Phi(x)\,\, \frac{\delta}{\delta \Phi(x)}\,\,,
\]
where $D$ is the space-time dimension,
fulfilling the same commutation rules as the charges:
\begin{equation}
\bigl[W_\alpha,W_\beta\bigr]=i f_{\alpha \beta \gamma} W_\gamma\,\,.
\label{comm-W}
\end{equation}
These operators act on $\Gamma[\F]$, the vertex functional which generates the 
1-particle irreducible and amputated Feynman graphs. In a perturbation expansion 
in powers of $\hbar$ -- equivalent to an expansion in the number of loops -- the zeroth 
order or tree approximation functional $\Gamma^{(0)}$ is just the classical action $S[\F]$, 
a local functional of the classical fields $\F$.

Now, suppose that, at the classical level, one  adds to an 
action $S_{\rm inv}[\F]$, invariant under the group transformations,
\ie  satisfying $W_\alpha S_{\rm inv}=0$, a {\it breaking term} 
\[
S_{\rm break}[\F] =  b^I\int d^Dx\, B_I(x)\,\,,
\]
where the $b^I$'s play the role of  ``coupling constants" 
and the $B_I$'s are local functionals of $\F$, \ie local polynomials 
in $\F$ and its derivatives. We assume that the $B_I$'s have power-counting 
dimension $d\leq D$, and transform under the symmetry transformations in a 
given representation $R$ of the group:
\[
 \d_\a B_I(x) = R_{\a I}{}^J B_J(x)\,\,,
\]
where the $R_{\a I}{}^J$ are representation matrix elements 
of the generators $Q_\a$. Then the total action 
\begin{equation}
S_{\rm tot}[\F] =  S_{\rm inv}[\F]+S_{\rm break}[\F]\,\,,
\label{Stot}
\end{equation} 
breaks the $G$ symmetry:
\begin{equation}
W_\a S_{\rm tot}=b^I R_{\a I}{}^J \int d^Dx\,  B_J \not= 0\,\,.
\label{WI-break}
\end{equation}
In order to {\it control the breaking} and, in particular 
its power-counting and symmetry
properties in all orders of perturbation theory, following 
Symanzik~\cite{Sym1,Sym2}, we convert this action into one 
which is invariant under the original transformation, adding  
a term involving external fields $\b_I(x)$ of power-counting 
dimension $D-d$:\footnote{This dimension will specify the renormalization 
procedure of the breaking operators $B_I$ defined below.}
\begin{equation}
S[\F,\b] = S_{\rm inv}[\F]+\int d^Dx\,(\b^I(x)+b^I)B_I(x)\,\,,
\label{new-action}
\end{equation}
where the $\b^I$'s transform under $G$ as
\[
\d_\a \b^I(x) = -(\b^J(x)+b^J) R_{\a I}{}^J B_J(x)\,\,.
\]
The new action is invariant:
 \begin{equation}
W_\a S[\F,\b] = -i\int d^Dx
\left(\d_\a \Phi(x) \frac{\d}{\d \Phi(x)} +
\d_\a \b^I(x) \frac{\d}{\d \b^I(x)} \right) S[\F,\b] = 0\,\,.
\label{WI-S}
\end{equation} 
Obviously, at $\b(x)=0$, $S$ reduces to the action (\ref{Stot}), 
and the identity (\ref{WI-S}) to the breaking identity (\ref{WI-break}).

The purpose of renormalization is to construct, perturbatively, 
a vertex functional $\G[\F,\b]$ obeying the same functional identity
(\ref{WI-S}), now expressed as the Ward identity
\begin{equation} 
W_\a \G[\F,\b] = 0\,\,.
\label{QWI}\end{equation}
This identity, taken at external field $\b=0$, yields the broken Ward identity
\begin{equation}
W_\a\G[\F,0]= b^J R_{\a,J}{}^I \int d^Dx\, 
\left.\frac{\d\G[\F,\b]}{\d \b^I(x)}\right|_{\b^I(x)=0}\,\,,
\label{QWI-break}\end{equation}
where the r.h.s. represents the renormalization of the classical breaking in the 
r.h.s. of (\ref{WI-break})~\cite{Sym1}-\cite{Stora2}.

At this point, two important remarks have to be done:
\begin{itemize}

\item[1.]
Since we have assigned to the external field $\b$ the power 
counting dimension $D-d$,
we know from the QAP that the dimension of the
renormalized breaking in the r.h.s. of (\ref{QWI-break}) is an insertion of 
dimension $d$. In particular, if $d<D$, it is guaranteed to be
a ``soft'' insertion, which means in particular that the asymptotic
behaviour in momentum space of a Green function with this insertion is lower 
by a power $D-d$ than the same green function without the insertion.

\item[2.] The broken Ward identity (\ref{QWI-break}) explicitly shows 
that the renormalized breaking belongs to the same representation
as its classical counterpart given in (\ref{WI-break}).

\end{itemize}

In conclusion, successfully fulfilling the renormalization program 
leading to the Ward identity (\ref{QWI}) yields a perturbative
quantization of the classical theory with full control of the 
dimension and covariance of the breaking. Of course, it remains the 
possibility of an anomaly, \ie the impossibility to fulfil the Ward identity.

The success of this program is guaranteed if the following two conditions are 
met~\cite{Sym1}-\cite{Stora3}:
\begin{itemize}

\item[1)] {\it Criterion of stability of the theory under small perturbations:} 
All possible counterterms 
$\D_{\rm c.t.}[\f,\b]$, solutions of the 
invariance conditions $W_\a\D_{\rm c.t.}=0$, correspond to the renormalization 
of the parameters and fields of the classical theory defined by the action (\ref{new-action}). 
If some solutions of the invariance condition do not meet this requirement, one has to 
suitably complete the classical action. As a consequence of the QAP, the  $\D_{\rm c.t.}$'s
are integrated local functionals of the fields $\F$ and $\b$, of dimension limited by $D$ due 
to power-counting renormalizability.

\item[2)] {\it Absence of anomaly:} Let $\D_\a[\F,\b]$ be integrated local functionals
of dimension up to $D$. Then, the {\it consistency conditions}
\begin{equation}
W_\a\D_\b - W_\b\D_\a = if_{\a\b\g} \D_\g\,\,,
\label{consistency}
\end{equation} 
admit only ``trivial'' solutions of the form
\begin{equation}
\D_\a = W_\a \D\,\,,
\label{triviality}
\end{equation}
for some integrated local functional $\D[\f,\b]$ of dimension at most 
equal to $D$. If this condition is not met, we say we have an anomaly.

\end{itemize}

Condition 2) is based on the fact that, as a consequence of 
the QAP, the possible breakings of the Ward identity (\ref{QWI}) are insertions of 
integrated local insertions whose lowest order are integrated local functionals
$\D_\a$ of dimension limited by $D$. The consistency condition (\ref{consistency})
are then a consequence of the algebra (\ref{comm-W}). The fulfilment of (\ref{triviality}) 
means that any possible breaking $\D_a$ can be reabsorbed in the action, at each 
order of perturbation theory, as non-invariant counterterm equal to $-\D$. As noticed 
first by Stora and his collaborators~\cite{Stora1,Stora2,Stora3}, solving this condition 
amounts to solving a problem of Lie algebra cohomology.

\section{\bf A toy model with hard Lorentz breaking}
\label{Sec3}
\hspace*{\parindent}
We study here a toy model of scalar fields with a 
{\it hard breaking} of Lorentz invariance, \ie a breaking of dimension 4,
in a 4D space-time with Minkovsky metric
$\eta_{\m\n}=\mbox{diag}(1,-1,-1,-1)$. The scalar fields form an $SO(N)$ multiplet $\vf^i(x)$,
$i=1,\cdots,N$,  transforming under the Lorentz transformations as
\[
\d\vf^i (x) = - \e^\m{}_\n x^\n\pa_\m \vf^i(x)\,\,,
\]
where $\e^{\m\n}=-\e^{\n\m}$ are infinitesimal parameters.
The invariant part of the action is the most general Lorentz invariant one, restricted by 
power-counting renormalizability to be
\begin{equation}
S_{\rm inv}[\vf] = \int d^4x\,\left(\frac12 \pa^\m\vf^i\pa_\m\vf^i 
-\frac{m^2}{2} \vf^i\vf^i - \frac{\lambda}{4!} (\vf^i\vf^i)^2 \right)\,\,,
\label{S-inv-toy}\end{equation}
with $m^2$ and $\lambda$ choosen as positive. The hard breaking term is given by
\[
S_{\rm break}[\vf] = \frac12 c^{\m\n}\int d^4x\,\pa_\m\vf^i\pa_\n\vf^i\,\,,
\]
where the 9 arbitrary numbers $c^{\m\n}$ are the elements of a symmetric traceless matrix 
(the trace part would correspond to a scalar term already present in $S_{\rm inv}$ as its kinetic term).

As explained in Section \ref{Sec2}, we need to introduce an external field in order to control the group 
theory characteristics of the breaking. The characteristics here is that of a symmetric, traceless Lorentz 
tensor of rank two:
\[
B_{\m\n} = \pa_\m\vf^i\pa_\n\vf^i - 
\frac14 \eta_{\m\n} \pa^\lambda\vf^i\pa_\lambda\vf^i\,\,.
\]
The external field coupled to this breaking will thus be a symmetric traceless tensor
field $ \g^{\m\n}(x)$, transforming as
\[
\d \g^{\m\n}(x)= - \e^\rho{}_\lambda x^\lambda\pa_\rho  \g^{\m\n}(x)
+ \e^\m{}_\rho (\g^{\rho\n}(x)+c^{\rho\n}) + \e^\n{}_\rho (\g^{\rho\m}(x)+c^{\rho\m})\,\,,
\]
under the Lorentz transformations.
The functional identity (\ref{WI-S}) takes here the form
\begin{equation}
W_{(\e)}S =  -i\int d^4x
\left(\d\vf^i(x) \frac{\d S}{\d \vf^i} +
\frac12\d  \g^{\m\n}(x) \frac{\d S}{\d  \g^{\m\n}(x)} \right) =0\,\,,
\label{WI-toy}\end{equation}
where the functional operators obey to the commutation rules
\begin{equation}
\left[ W_{(\e)},W_{(\eta)} \right] = i W_{([\e,\eta])\,}\,,
\label{comm-toy}\end{equation}
An action solution of (\ref{WI-toy}) is
\begin{equation}
S = S_{\rm inv} + 
\frac12\int d^4x\,(\g^{\m\n}+c^{\m\n})\pa_\m\vf^i\pa_\n\vf^i\,\,.
\label{action-toy}\end{equation}
In order to see if the theory thus defined is renormalizable, let us 
look to the criteria enumerated at the end of Section \ref{Sec2}.

\subsubsection*{Condition of stability:}
\hspace*{\parindent}
The solutions of the condition $W_{(\e)}\D_{\rm c.t.}$ = 0 are, 
either any one of the three terms of the invariant action 
(\ref{S-inv-toy}), or  $\g$-dependent terms, given by:
\begin{eqnarray}
\label{ct-toy1}
&\displaystyle{\sum_{n=1}^\infty}\displaystyle{ \frac{\a_n}{2n!}} \displaystyle{\int} d^4x\,\,
\hat\g^{\m}{}_{\rho_1}\hat\g^{\rho_1}{}_{\rho_2}\cdots\hat\g^{\rho_n\n}
\,\pa_\m\vf^i\pa_\n\vf^i\,\,,\\[3mm]
\label{ct-toy2}
&\displaystyle{\sum_{n=1}^\infty}\displaystyle{ \frac{\b_n}{2n!}} \displaystyle{\int} d^4x\,\,
\hat\g^{\m}{}_{\rho_1}\hat\g^{\rho_1}{}_{\rho_2}\cdots\hat\g^{\rho_n\n}
\,\pa_\m(\vf^i\pa_\n\vf^i)\,\,,
\end{eqnarray}
where $\hat\g^{\m\n}(x)=\g^{\m\n}(x)+c^{\m\n}$ and $\a_n$, $\b_n$ are arbitrary parameters. 
This set of counterterms is infinite due to the zero dimensionality of the exterior field $\g$. 
But one observes that the sum of terms in (\ref{ct-toy1}) can be reduced to the 
form of the $(\g+c)$ term of (\ref{action-toy}) by a non-linear redefinition of $\g$. 
However the counterterms in the sum (\ref{ct-toy2}) do not correspond to anything  
present in the action (\ref{action-toy}), and thus the latter should be completed 
with them -- although their role is trivial: they turn out to be total derivatives when 
setting $\g$ = 0 at the end. In summary, all possible  counterterms correspond to 
the renormalization of the given classical action: the parametrs $m$ and 
$\lambda$,  the field $\vf^i$ and the external field  $\g^{\m\n}$.
Let us note that the renormalization of $\g^{\m\n}$ amounts to a 
renormalization of the composite breaking operator $B_{\m\n}$.

\subsubsection*{Absence of anomalies:}
\hspace*{\parindent}
The form of the consistency condition follows from the commutation 
rules (\ref{comm-toy}) for (rigid) Lorentz symmetry:
\[
W_{(\e)} \D_{(\eta)} - W_{(\eta)} \D_{(\e)} =  \D_{([\e,\eta])}\,\,.
\]
Its general solution has been proved~\cite{Balasin} to be of the
form $\D_{\e}$ = $ W_{\e}\D$ with $\D$ an integrated local functional of dimension $\le4$. 
There is thus no anomaly.

\subsubsection*{Counting the number of parameters and of renormalizations:}
\hspace*{\parindent}
As we have seen in the discussion of stability under small perturbations, 
the theory depends on the following physical parameters: the mass $m$, the coupling
constant $\lambda$ and the 9 breaking parameters $c^{\m\n}$. Beyond these, 
the theory also depends on the parameter corresponding to the renormalization of the field
$\vf$ and on the infinite number of parameters $\a_n$ corresponding to the non-linear 
renormalization of the external field $\g^{\m\n}$, see (\ref{ct-toy1}).
On the other hand, since the  $c^{\m\n}$ are given in the definition of the
Ward identity operator $W_{\e}$, they are not renormalized. Thus we have only 2 
renormalizations, those of $m$ and $\lambda$, beyond the field amplitude renormalizations 
$\vf^i$ and $\g^{\m\n}$.

This situation has to be contrasted with that of  the theory with Lorentz invariance
hardly broken without control through external fields, as we are going to show later on.

Before going on, a remark concerning the infinity of parameters $\a_n$ is due. 
This seems to imply a non-renormalizability of the theory, since an infinite number 
of normalization conditions is indeed needed in order to fix them. 
The normalization conditions, beyond the usual ones which fix $m$, $\lambda$ and the field
amplitude of $\vf$, may be choosen, \eg as 
\[
\G_{2,n} = 1\,,\quad n=1,\cdots,\infty\,\,,
\]
where $\G_{2,n}$ is the vertex function with $n$ insertion of 
the breaking operator $\int d^4x\,c^{\m\n}B_{\m\n}$ 
and 2 (amputated) external lines $\vf^i$ at some conveniently choosen momentum.
Each of the counterterms in the sum (\ref{ct-toy1})
will contribute to the breaking, at zero external field, with
a coefficient of order $n$ in  $c^{\m\n}$. Therefore, 
in a realistic situation where the breaking is expected to be very small, only a few 
terms of low order will effectively contribute! 
  
\subsubsection*{Hard breaking without controlled covariance:}
\hspace*{\parindent}
Let us now turn to the case of a breaking of Lorentz symmetry where we do not introduce 
the controlling external field $\g^{\m\n}$, as, \eg in~\cite{Nos2}.
The reason for introducing external fields is two-fold, as we explain in Section \ref{Sec2}:
controlling the power-counting dimension of the breaking and controlling its covariance.
In the case  considered here,  no control of dimension is needed since 
the dimension of a hard breaking is by definition already the maximum one allowed by 
power-counting renormalizability, namely 4. 
In order to clarify some point which may still appear unclear, 
it is interesting to have a look on what would happen
if no controlling external field is introduced. We would thus start with the action 
(\ref{action-toy}) with $\g^{\m\n}=0$:
\begin{equation}
S = S_{\rm inv} + c^{\m\n}\int d^4x\,\pa_\m\vf^i\pa_\n\vf^i\,\,.
\label{action-hard}
\end{equation}
There is now no Ward identity like (\ref{WI-toy}), so that the
independent counterms which will be generated are all possible field polynomial of 
power-counting dimension up to 4, restricted only by $SO(N)$ invariance. Beyond the 
12 terms of the action (\ref{action-hard}), there is the possibility of the dimension 3 terms
$d^\m\!\!\int d^4x\,\vf^i\pa_\m\vf^i$ depending on 4 parameters $d^\m$. This total of 16 
counterterms do correspond to an equal number of 16 renormalizations.

How would one explain this difference in a context of Feynman graphs calculations?
On the one hand, calculating Green functions with the Feynman rules defined by the action 
(\ref{action-hard}), one will sooner or later encounter ultraviolet singularities whose 
renormalization will generate  16 arbitrary finite parameters corresponding to the 16 
ambiguities of the subtraction ~\cite{Zimmermann} or regularization~\cite{Epstein-Glaser} 
procedure.\footnote{For simplification, this procedure is assumed to preserve the explicit 
$SO(N)$ symmetry.} On the other hand, if one introduces the external fields and takes into 
account, at each order of perturbation theory, of the conditions imposed by the fulfilment 
of the Ward identity, the  dimension 3 terms  $d^\m\!\!\int d^4x\,\vf^i\pa_\m\vf^i$ will not 
appear and the independent parameters will be reduced to those we had above -- without 
counting the parameters $\a_n$ corresponding to the renormalization of the breaking operator.  

\section{\bf The QED with soft Lorentz breaking}
\label{Sec4}
\hspace*{\parindent}
The quantum electrodynamics (QED) with violation of Lorentz and CPT have been studied
intensively in recent years. Among several issues, the possible generation of a
Chern-Simons-like term induced by radiative corrections arising from a CPT and Lorentz
violating term in the fermionic sector has been a recurrent theme in the literature.
We particularly mention the following works~\cite{Ref1}-\cite{Mariz} (and references
cited therein), where many controversies have emerged from the discussion whether this
Chern-Simons-like term could be generated by means of radiative corrections arising
from the axial coupling of charged fermions to a constant vector $b_\mu$ responsible
for the breakdown of Lorentz Symmetry.

In this section, we reassess the discussion on the radiative generation of a
Chern-Simons-like term induced from quantum corrections in the extended QED.
We show, to all orders in perturbation theory, that a CPT-odd and Lorentz violating Chern-Simons-like term, 
{\it definitively}, is not radiatively induced by the axial coupling of the fermions with the constant vector $b_\mu$.
The proof of this fact is based on general theorems of perturbative quantum field theory 
(see~\cite{piguet-rouet,PigSor,Man} and references there in), where the Lowenstein-Zimmermann 
subtraction scheme in the framework of Bogoliubov-Parasiuk-Hepp-Zimmermann-Lowenstein 
(BPHZL) renormalization method \cite{Low} is adopted. The former has to be introduced, owing to the presence 
of massless gauge field, so as to subtract infrared divergences that should arise from the ultraviolet subtractions.

\subsection{The model at the classical level}
\hspace*{\parindent}
We start by considering an action for extended QED with a term which violates the
Lorentz and CPT symmetries in the matter sector only. In the tree approximation,
the classical action of extended QED with one Dirac spinor that we are considering
here is given by:
\begin{equation}
\Sigma^{(s-1)}=\Sigma_{\rm S}+\Sigma_{\rm SB}+\Sigma_{\rm IR}
+\Sigma_{\rm gf}+\Sigma_{\rm ext}\,\,,
\label{1}
\end{equation}
where
\begin{equation*}
\Sigma_{\rm S}=
\int d^{4}x\,\,\Big\{i\bar{\psi}\gamma^{\mu}(\partial_{\mu}+ieA_{\mu})\psi -
m \bar{\psi}\psi-\frac{1}{4}F^{\mu\nu}F_{\mu\nu}\Big\}\,\,,
\end{equation*}
is the symmetric part of $\Sigma$ under gauge and Lorentz transformations.
The term
\begin{equation}
\Sigma_{\rm SB}=-\int d^{4}x\,\,b_{\mu}\bar{\psi}\gamma_{5}\gamma^{\mu}\psi\,\,, 
\label{SB}
\end{equation}
is the symmetry-breaking part of $\Sigma$ that breaks the manifest Lorentz
covariance by the presence of a constant vector $b_{\mu}$ which selects a
preferential direction in Minkowski space-time, breaking its isotropy, as
well as it breaks CPT. In turn,
\begin{equation*}
\Sigma_{\rm IR}=\int d^{4}x\,\,\frac{1}{2}M^2 (s-1) A_\mu A^\mu\,\,,
\end{equation*}
is the Lowenstein-Zim\-mer\-mann mass term for the photon field. A Lowenstein-Zim\-mer\-mann mass term 
of the $M^2(s-1)$ type is there in order to enable a momentum space subraction scheme without 
introducing spurious infrared singularities. The Lowenstein-Zimmermann parameter $s$ lies in the interval 
$0 \leq s \leq 1$ and plays the role of an additional subtraction variable (as the external momentum) in the 
BPHZL renormalization program, such that the theory describing a really massless particle is
recovered for $s=1$. At this point, a comment about the Lowenstein-Zimmermann mass term for the photon 
field is now in order: the gauge invariance properties are not spoiled by the photon mass; this is a peculiarity 
of the {\it abelian case} \cite{piguet-rouet}. This was studied in details for the QED in Ref.\cite{low-schroer} 
using the BPHZ scheme.

Finally, in order to quantize the system a gauge-fixing is added 
\begin{equation}
\Sigma_{\rm gf}=\int d^{4}x~\left(b \partial_\mu A^\mu+\frac{\xi}{2}b^2 
+ \overline{c} \square c\right)~,
\label{gf}
\end{equation}
together with the term, $\Sigma_{\rm ext}$, coupling the non-linear Becchi-Rouet-Stora-Tyutin 
(BRST) transformations to external sources
\begin{equation}
\Sigma_{\rm ext}=\int d^{4}x~\left(\overline{\Omega}{\mathfrak s}
\psi-{\mathfrak s}{\overline \psi}\Omega\right)~,
\label{ext}
\end{equation}

\subsubsection*{Continuous symmetries:}
\hspace*{\parindent}
The infinitesimal BRST transformations are given by:
\begin{eqnarray}
&{\mathfrak s} \psi=ic \psi~,~~{\mathfrak s} \overline{\psi}=-ic \overline \psi~,\nonumber \\[3mm]
&{\mathfrak s} A_\mu=-\frac{1}{e} \partial_\mu c~,~~ {\mathfrak s} c=0~, \label{BRS} \\[3mm]
&{\mathfrak s} {\overline c}=\frac{1}{e}b~~,~~{\mathfrak s} b=0~, \nonumber
\end{eqnarray}
where $c$ is the ghost field, ${\overline c}$ is the antighost field and $b$ is the Lautrup-Nakanishi 
field~\cite{lautrup-nakanishi}, respectively. The latter plays the role of the Lagrange multiplier field. 
Although not massive, the Faddeev-Popov ghosts are free fields, they decouple, therefore, 
no Lowenstein-Zimmermann mass term has to be introduced for them.

The BRST invariance of the action is expressed in a functional way by the Slavnov-Taylor identity
\begin{equation}
\cs(\S^{(s-1)})=0\,\,,
\label{slavnovident}
\end{equation}
where the Slavnov-Taylor operator $\cs$ is defined, acting on an arbitrary functional $\cf$, by
\begin{equation}
\cs(\cf)=\int{d^4 x} \biggl\{-{1\over e}{\pa}^\mu c {\d\cf\over\d A^\mu} + {1\over e}b {\d\cf\over\d {\ov c}} + 
{\d\cf\over\d \ov\O}{\d\cf\over\d \j} - {\d\cf\over\d \O}{\d\cf\over\d \ov\j}\biggl\}\,\,.
\label{slavnov}
\end{equation}
The corresponding linearized Slavnov-Taylor operator reads
\begin{equation}
\cs_\cf=\int{d^4 x} \biggl\{-{1\over e}{\pa}^\mu c {\d\over\d A^\mu} + {1\over e}b {\d\over\d {\ov c}} 
+{\d\cf\over\d \ov\O}{\d\over\d \j} + {\d\cf\over\d \j}{\d\over\d \ov\O} 
- {\d\cf\over\d \O}{\d\over\d \ov\j} - {\d\cf\over\d\ov\j}{\d\over\d \O} \biggl\}\,\,.
\label{slavnovlin}
\end{equation}
The following nilpotency identities hold:
\begin{align}
\cs_\cf\cs(\cf)&=0~,~~\forall\cf~, \label{nilpot1} \\[3mm]
\cs_\cf\cs_\cf&=0~~{\mbox{if}}~~\cs(\cf)=0~. 
\label{nilpot3}
\end{align}
In particular, $(\cs_\S^{(s-1)})^2=0$, since the action $\S^{(s-1)}$ obeys
the Slavnov-Taylor identity (\ref{slavnovident}). The operation of
$\cs_{\S^{(s-1)}}$ upon the fields and the external sources is given by
\begin{align*}
&\quad\cs_{\S^{(s-1)}}\f=s\f~,~~\f=\{\j,\ov\j,A_\m,c,{\ov c},b\}~, \\[3mm]
&\cs_{\S^{(s-1)}}\O=-{\d\S^{(s-1)}\over\d\ov\j}~,~~
\cs_{\S^{(s-1)}}\ov\O_+={\d\S^{(s-1)}\over\d\j}\,\,. 
\end{align*}

In addition to the Slavnov-Taylor identity (\ref{slavnovident}), the classical action 
$\S^{(s-1)}$ (\ref{1}) is characterized by the gauge condition, the ghost equation and the
antighost equation:
\begin{align}
{\d\S^{(s-1)}\over\d b}&=\pa^\m A_\m + \x b~,\label{gaugecond} \\[3mm]
{\d\S^{(s-1)}\over\d \ov c}&=\square c~,\label{ghostcond} \\[3mm]
-i{\d\S^{(s-1)}\over\d c}&=i\square{\ov c} + \ov\O\j - \ov\j\O~. 
\label{antighostcond}
\end{align}

The action $\S^{(s-1)}$ (\ref{1}) is invariant also with respect to the rigid symmetry
\begin{equation}
W_{\rm rigid} \S^{(s-1)}=0~, 
\label{rigidcond}
\end{equation}
where the Ward operator, $W_{\rm rigid}$, is defined by
\begin{equation*}
W_{\rm rigid}=\int{d^4 x}\biggl\{\j{\d\over\d \j} - \ov\j{\d\over\d \ov\j} 
+ \O{\d\over\d \O} - \ov\O{\d\over\d \ov\O} \biggr\}~. 
\end{equation*}

On the other hand, the Lorentz symmetry is broken by the presence of the 
constant vector $b_\m$. The fields $A_\m$ and $\j$ transform under
infinitesimal Lorentz transformations $\d x^\m$ =$\e^\m{}_\n x^\n$,
with $\e_{\m\n}=-\e_{\n\m}$, as
\begin{eqnarray}
&\delta_{\rm L} A_\mu =-\e^\l{}_\n x^\n\pa_\l A_\m + \e_\m{}^\n A_\n
\equiv \frac12 \e^{\a\b} \d_{\rm L\a\b} A_\m\,\,, \nonumber\\[3mm]
&\delta_{\rm L}\j  =-\e^\l{}_\n x^\n\pa_\l \j
-\frac{i}{4}\e^{\m\n}\s_{\m\n}\j 
\equiv \frac12 \e^{\a\b} \d_{\rm L\a\b} \j\,\,, 
\label{lor-var-A-psi}
\end{eqnarray} 
where $\sigma_{\mu\nu}=\frac{i}{2}\,[\gamma_\mu,\gamma_\nu]$.

It should be noticed that the Lorentz breaking (\ref{SB}) 
is not linear in the dynamical fields, therefore will be renormalized.
It is however a ``{\it soft breaking},'' since its UV power-counting
dimension is less than 4, namely 3. 
According to Symanzik~\cite{Sym1,Sym2}, a theory with soft
symmetry breaking is renormalizable if the radiative corrections do not
induce a breakdown of the symmetry by terms of UV power-counting
dimension equal to 4 -- called hard breaking terms. Concretely, according to 
the Weinberg's Theorem~\cite{weinberg}, this means that the symmetry of the 
theory in the asymptotic deep euclidean region of momentum space is preserved 
by the radiative corrections. In order to control the Lorentz breaking and, in 
particular, its power-counting properties, following Symanzik~\cite{Sym1,Sym2}, 
and~\cite{Balasin} for the specific case of Lorentz breaking, we introduce
an external field $\b_\m(x)$, of dimension 1 and transforming under Lorentz
transformations according to
\begin{equation}
\delta_{\rm L}\b_\mu(x)=-\e^\l{}_\n x^\n\pa_\l \b_\m(x)+
\e_\m{}^\n (\b_\n(x)+b_\n)
\equiv \frac12 \e^{\a\b} \d_{\rm L\a\b} \b_\m(x)\,\,.
\label{lor-var-beta}
\end{equation}
The functional operator which generates these transformations reads
\begin{equation}
{W}_{\rm L\a\b}=\int d^4x~{W}_{\rm L\a\b}(x)=
\int d^4x\,\sum_{\vf=A_\m,\j,\bar\j,\b}\d_{\rm L\a\b}\vf(x)\frac{\d}{\d \vf(x)}\,\,.
\label{Lor-w-op}
\end{equation} 
Redefining the action by adding a term in $\b_\mu$:
\begin{equation}
\widetilde\S^{(s-1)} = \S^{(s-1)} - \int d^4x\,\b_\m\bar\j\g_5\g^\m\j\,\,,
\label{25}
\end{equation}
one easily checks the classical Ward identity
\begin{equation}
{W}_{\rm L\a\b}\widetilde\Sigma^{(s-1)}=0\,\,,
\label{tilde-Lor-Ward}
\end{equation} 
which, at $\b_\m=0$, reduces to the broken Lorentz Ward identity
\begin{equation}
{W}_{\rm L\a\b}\Sigma^{(s-1)} = 
\e_\m{}^\n b_\n\int d^4x\,\bar\j\g_5\g^\m\j\,\,.\label{Lor-Ward}
\end{equation} 
The external field $\beta_\mu(x)$ being coupled to
a gauge invariant expression (the axial current: 
$j_5^\mu=\bar{\psi}\gamma_{5}\gamma^{\mu}\psi$), 
we take it to be BRST invariant in order to preserve gauge invariance,
\begin{equation}
{\mathfrak s} \int d^4x\,\b_\m\bar\j\g_5\g^\m\j
=0 ~\Longrightarrow~ {\mathfrak s}\beta_\mu(x)=0\,\,.
\end{equation} 
Therefore, it follows that the action $\widetilde{\Sigma}^{(s-1)}$ (\ref{25}) 
satisfies the same Slavnov identity 
(\ref{slavnovident}) as the action $\Sigma^{(s-1)}$ (\ref{1}), namely: 
\begin{equation}
\cs(\widetilde{\Sigma}^{(s-1)})=0\,\,,
\label{tilde-slavnov}
\end{equation}
together with the conditions:
\begin{align}
{\d\widetilde\S^{(s-1)}\over\d b}&=\pa^\m A_\m + \x b~,\label{tilde-gaugecond} \\[3mm]
{\d\widetilde\S^{(s-1)}\over\d \ov c}&=\square c~,\label{tilde-ghostcond} \\[3mm]
-i{\d\widetilde\S^{(s-1)}\over\d c}&=i\square{\ov c} + \ov\O\j - \ov\j\O~, \label{tilde-antighostcond} \\[3mm]
W_{\rm rigid} \widetilde\S^{(s-1)}&=0~, \label{tilde-rigidcond}
\end{align}

\subsubsection*{Discrete symmetries:}
\hspace*{\parindent}
The discrete symmetries of the theory are the following ones:

\,\,\,\noindent {\it Charge conjugation} $C$:
Assuming the Dirac representation of the $\gamma$-matrices \cite{itzykson}, the 
charge conjugation transformations read: 
\begin{align}\label{9}
\psi \stackrel{C}{\longrightarrow} \psi^c &=C\,{\bar{\psi}}^T\,\,,\nonumber \\
\bar{\psi} \stackrel{C}{\longrightarrow} \bar{\psi}^c &=-\psi^T C^{-1}\,\,,\nonumber \\
A_\mu \stackrel{C}{\longrightarrow} A_{\mu}^{c} &=-A_{\mu}\,\,, \nonumber \\
C\gamma_{\mu}C &= \gamma_{\mu}^{T}\,\,,\nonumber \\
C\gamma_{5}C &=-\gamma_{5}^{T}=-\gamma_{5}\,\,.
\end{align}
where $C$ is the charge conjugation matrix, with $C^{2}=-1$. 
All terms of the action $\widetilde{\Sigma}^{(s-1)}$ (\ref{25}) are invariant under
charge conjugation.

\noindent {\it Parity} $P$:
\begin{align}\label{11}
x &\stackrel{P}{\longrightarrow} (x^0,\,-\vec{x})\,\,, \nonumber\\
\psi &\stackrel{P}{\longrightarrow} \gamma^{0}\psi\,\,, \nonumber\\
\bar{\psi} &\stackrel{P}{\longrightarrow} \bar{\psi}\gamma^{0}\,\,, \nonumber\\
A_{\mu} &\stackrel{P}{\longrightarrow} A^{\mu}\,\,. 
\end{align}
All terms of the action $\widetilde{\Sigma}^{(s-1)}$ (\ref{25}) are invariant under parity, 
unless the Lorentz breaking term $\Sigma_{\rm SB}$ (\ref{SB}).

\noindent {\it Time reversal} $T$:
\begin{align}\label{13}
\psi &\stackrel{T}{\longrightarrow} T\psi\,\,,\nonumber \\
\bar{\psi} &\stackrel{T}{\longrightarrow} \bar{\psi}T\,\,,\nonumber \\
A_{\mu} &\stackrel{T}{\longrightarrow} A_{\mu}\,\,, \nonumber \\
T\gamma^{\mu}T &= \gamma_{\mu}^{T}=\gamma^{\mu*}\,\,,\nonumber \\
T\gamma_{5}T &= \gamma_{5}\,\,.
\end{align}
The broken Lorentz term $\Sigma_{\rm SB}$ (\ref{SB}) are non invariant under time reversal, 
whereas the other terms in the action $\widetilde{\Sigma}^{(s-1)}$ (\ref{25}) remain invariant. 
As a consequence, the action $\widetilde{\Sigma}^{(s-1)}$ (\ref{25}), has CPT symmetry 
broken by the Lorentz breaking term, $\Sigma_{\rm SB}$ (\ref{SB}): 
\begin{equation}\label{15}
\bar{\psi}b_{\mu}\gamma_{5}\gamma^{\mu}\psi
\stackrel{CPT}{\longrightarrow}
-\bar{\psi}b_{\mu}\gamma_{5}\gamma^{\mu}\psi\,\,.
\end{equation}

\subsubsection*{UV and IR dimensions:}
\hspace*{\parindent}
By switching off the coupling constant ($e$) and taking the free part of the action (\ref{1}), 
the tree-level propagators in momenta space, for all the fields, read: 
\begin{align}
&\Delta_{\psi\psi}(k)=i\frac{\sl{k}+m}{k^2-m^2}~,\label{propk++--} \\[3mm]
&\Delta^{\mu\nu}_{AA}(k,s)=-i\biggl\{\frac{1}{k^2-M^2(s-1)^2}\biggl(\eta^{\mu\nu}-\frac{k^\mu k^\nu}{k^2}\biggr)
+ \frac{\xi}{k^2-\xi\,M^2(s-1)^2}\frac{k^\mu k^\nu}{k^2} \biggr\}~, \label{propkAA} \\[3mm]
&\Delta^\mu_{Ab}(k)=\frac{k^\mu}{k^2}~,~~\Delta_{bb}(k)=0~, \label{propkAbbb} \\[3mm]
&\Delta_{{\overline c}c}(k)=-i\frac{1}{k^2}~.\label{propkcbc}
\end{align}

In order to establish the ultraviolet (UV) and infrared (IR) dimensions of any fields, 
$X$ and $Y$, we make use of the UV and IR asymptotical behaviour of their propagator, 
$\Delta_{XY}(k,s)$, $d_{XY}$ and $r_{XY}$, respectively: 
\begin{align*}
d_{XY}={\overline{\rm deg}}_{(k,s)}\Delta_{XY}(k,s)\,\,, \\[3mm] 
r_{XY}={\underline{\rm deg}}_{(k,s-1)}\Delta_{XY}(k,s)\,\,,
\end{align*}
where the upper degree ${\overline{\rm deg}}_{(k,s)}$ gives the asymptotic power 
for $(k,s)\rightarrow \infty$ whereas the lower degree ${\underline{\rm deg}}_{(k,s-1)}$ 
gives the asymptotic power for $(k,s-1) \rightarrow 0$. The UV ($d$) and IR ($r$) 
dimensions of the fields, $X$ and $Y$, are chosen to fulfill the following inequalities:
\begin{align}
d_X + d_Y \geqslant 4 + d_{XY} \quad {\mbox{and}} \quad r_X + r_Y \leqslant 4 + r_{XY}\,\,. 
\label{uv-ir}
\end{align}

In summary, the UV ($d$) and IR ($r$) dimensions -- which are those involved in the Lowenstein-Zimmermann 
subtraction scheme \cite{Low} -- as well as the ghost numbers ($\Phi\Pi$) and the Grassmann parity (GP) of 
all fields are collected in Table \ref{table1}. Notice that the statistics is defined as follows: the integer spin fields 
with odd ghost number, as well as, the half integer spin fields with even ghost number anticommute among themselves. 
However, the other fields commute with the formers and also among themselves.

\begin{table}
\begin{center}
\begin{tabular}{|c||c|c|c|c|c|c|c|c|c|}
\hline
    &$A_\mu$ &$\psi$ &$c$ &${\overline c}$ &$b$ &$\Omega$ &$\beta_\mu$ &$s-1$ &$s$ \\
\hline\hline
$d$ &1 &${3/2}$ &0 &2 &2 &5/2 &1 &1 &1 \\
\hline
$r$ &1 &2 &0 &2 &2 &2 &1 &1 &0 \\
\hline
$\Phi\Pi$&0 &0 &1 &$-1$ &0 &$-1$ &0 &0 &0 \\
\hline
$GP$&0 &1 &1 &1 &0 &0 &0 &0 &0 \\
\hline
\end{tabular}
\end{center}
\caption[]{UV ($d$) and IR ($r$) dimensions, ghost number ($\Phi\Pi$) and Grassmann parity ($GP$).}
\label{table1}
\end{table}

\subsection{The model at the quantum level}
\hspace*{\parindent}
In this section, we present the perturbative quantization of the extended QED
theory, using the algebraic renormalization procedure (see~\cite{PigSor,Man,PigDan} for
a review of the method and references to the original literature).
Our aim is to prove that the full quantum theory has the same properties
as the classical theory, {\it i.e.}, demonstrate that, at the quantum level, the Slavnov-Taylor 
identity, related to the gauge symmetry (\ref{tilde-slavnov}), and the Ward identity associated 
to the Lorentz symmetry (\ref{tilde-Lor-Ward}), are satisfied to all orders of perturbation theory:
\begin{eqnarray}
&&\cs({\G}^{(s-1)})|_{s=1}=0~, \label{quantum-slavnov} \\[3mm]
&&{W}_{\rm L\a\b}\G^{(s-1)}|_{s=1}=0~.\label{quantum-Lor-Ward}
\end{eqnarray}
In order to study the renormalizability of models characterized by a system of Ward 
identities, without referring to any special regularization procedure, two steps
must be followed~\cite{PigSor,Man,PigDan}: In the first step, we compute the possible
anomalies  of the Ward identities through an analysis of the Wess-Zumino
consistency condition. Next, we  check the stability of the classical action -- which ensures 
that the quantum corrections do not produce counterterms corresponding to the renormalization 
of parameters not already present in the classical theory.

\subsection{The Wess-Zumino consistency condition: in search for anomalies}
\hspace*{\parindent}
At the quantum level the vertex functional, $\G^{(s-1)}$, which coincides with the classical action, 
$\widetilde\S^{(s-1)}$ (\ref{25}), at $0$th order in $\hbar$,
\begin{equation}
\G^{(s-1)}=\widetilde\S^{(s-1)} + {\co}(\hbar)~,\label{vertex}
\end{equation}
has to satisfy the same constraints as the classical action does, namely  
Eq.(\ref{tilde-Lor-Ward}) and Eqs.(\ref{tilde-slavnov})-(\ref{tilde-rigidcond}).

In accordance with the Quantum Action Principle \cite{Lam,Lowenstein,Brenneke-Dutsch}, 
the Slavnov-Taylor identity (\ref{slavnovident}) and the Lorentz symmetry Ward identity get 
a quantum breakings:
\begin{eqnarray}
&&\cs(\G^{(s-1)})|_{s=1}=\D \cdot \G^{(s-1)}|_{s=1} = \D_{\rm g} + {\co}(\hbar \D_{\rm g})~,\label{slavnovbreak}\\[3mm]
&&{W}_{\rm L\a\b}\G^{(s-1)}|_{s=1}=\D_{\rm L\a\b} \cdot \G^{(s-1)}|_{s=1} = \D_{\rm L\a\b} + {\co}(\hbar \D_{\rm L\a\b})~,
\label{lorentzbreak}
\end{eqnarray}
where $\D_{\rm g}\equiv\D_{\rm g}|_{s=1}$ and $\D_{\rm L\a\b}\equiv\D_{\rm L\a\b}|_{s=1}$ are 
integrated local functionals, taken at $s=1$, with ghost number one and, UV and IR dimensions 
bounded by $d\le4$ and $r\ge4$, respectively.

The validity of the Lorentz Ward identity has been proved in~\cite{Balasin} by using the Whitehead's Lemma 
for semi-simple Lie groups, which states the vanishing of the first cohomology of such kind of group~\cite{Stora1,Stora3}. 
Here, see details in~\cite{Nos1}, this means that $\D_{\rm L\a\b}$ in (\ref{lorentzbreak}) can be written as
\begin{equation}
\Delta_{{\rm L}\a\b}={W}_{{\rm L}\a\b}\widehat\Delta_{\rm L}~, \label{DeltaL}
\end{equation}
where $\widehat\Delta_{\rm L}$ is an integrated local insertion of UV and IR dimensions 
bounded by $d\le4$ and $r\ge4$, respectively. Therefore, $\widehat\Delta_{\rm L}$ can be
reabsorbed in the action as a noninvariant counterterm, order by order, establishing the Lorentz Ward 
identity (\ref{quantum-Lor-Ward}) at the quantum level.

The nilpotency identity ({\ref{nilpot1}) together with
\begin{equation}
\cs_{\G^{(s-1)}}=\cs_{\widetilde\S^{(s-1)}} + {\co}(\hbar)~,
\end{equation}
implies the following consistency conditions for the breaking $\D_{\rm g}$:
\begin{equation}
\cs_{\widetilde\S^{(s-1)}}\D_{\rm g}=0~,\label{breakcond1}
\end{equation}
and beyond that, the breaking $\D_{\rm g}$ also satisfy the constraints:
\begin{equation}
{\d\D_{\rm g}\over\d b}={\d\D_{\rm g}\over\d\ov c}=\int d^4x \frac{\d\D_{\rm g}}{\d c}=
W_{\rm rigid}\D_{\rm g}={W}_{\rm L\a\b}\D_{\rm g}=0~.\label{breakcond5}
\end{equation}

The Wess-Zumino consistency condition (\ref{breakcond1}) constitutes a
cohomology problem in the sector of ghost number one.
Its solution can always be written as a sum of a trivial cocycle
$\cs_{\S}{\wh\D_{\rm g}}^{(0)}$, where ${\wh\D_{\rm g}}^{(0)}$ has ghost number zero,
and of nontrivial elements belonging to the cohomology of $\cs_{\widetilde\S^{(s-1)}}$
(\ref{slavnovlin}) in the sector of ghost number one:
\begin{equation}
\D_{\rm g}^{(1)} = {\wh\D_{\rm g}}^{(1)} + \cs_{\widetilde\S^{(s-1)}}{\wh\D_{\rm g}}^{(0)}~.\label{breaksplit}
\end{equation}
However, considering the Slavnov-Taylor operator $\cs_{\widetilde\S^{(s-1)}}$ (\ref{slavnovlin}) and the 
quantum breaking (\ref{slavnovbreak}), it results that $\D_{\rm g}^{(1)}$ exhibits UV and IR dimensions 
bounded by $d\leq4$ and $r\geq4$.

From the antighost equation in (\ref{breakcond5}): 
\begin{equation}
\int d^4x \frac{\d\wh\D_{\rm g}^{(1)}}{\d c}=0~,
\end{equation}
it follows that $\wh\D_{\rm g}^{(1)}$ can be written as 
\begin{equation}
\wh\D_{\rm g}^{(1)} = \int{d^4 x}~{\cal T}_\m\pa^\m c~, \label{delta-T}
\end{equation}
where ${\cal T}_\mu$ is a rank-$1$ tensor with ghost number zero, with UV and IR dimensions 
bounded by $d\leq3$ and $r\geq3$, respectively. The tensor ${\cal T}_\mu$ can be split into two 
pieces: 
\begin{equation}
{\cal T}_\mu = r_{\rm v} {\cal V}_\mu + r_{\rm p} {\cal P}_\mu~, \label{T-tensor}
\end{equation}
where ${\cal V}_\mu$ is a vector and ${\cal P}_\mu$ is a pseudo-vector, with $r_{\rm v}$ and $r_{\rm p}$ 
being coefficients to be determined. By considering the UV and IR dimensional constraints to be satisfied 
by ${\cal T}_\mu$ (\ref{T-tensor}) together with the conditions upon the Slavnov-Taylor breaking 
$\wh\D_{\rm g}^{(1)}$ (\ref{breaksplit}), given by (\ref{breakcond1}) and (\ref{breakcond5}), it follows that:
\begin{equation}
{\cal T}_\mu = r_{\rm v} \pa^\r F_{\r\m} + r_{\rm p} \epsilon_{\m\n\r\s}A^\n F^{\r\s} ~. \label{T-tensorf}
\end{equation}
Consequently, substituting (\ref{T-tensorf}) into (\ref{delta-T}), the breaking $\wh\D_{\rm g}^{(1)}$ reads:
\begin{equation}
\wh\D_{\rm g}^{(1)} = -\frac{r_{\rm p}}{2} \int{d^4 x}~c\epsilon_{\m\n\r\s}F^{\m\n}F^{\r\s} ~, \label{ABBJ}
\end{equation} 
which is the (Abelian) Adler-Bardeen-Bell-Jackiw anomaly \cite{abbj}. Therefore, up to noninvariant 
counterterms, which are $\cs_{\widetilde\S^{(s-1)}}$-variations of the integrated local insertions 
${\wh\D_{\rm g}}^{(0)}$:
\begin{equation}
\D_{\rm g}^{(1)} = \cs_{\widetilde\S^{(s-1)}}{\wh\D_{\rm g}}^{(0)}
-\frac{r_{\rm p}}{2} \int{d^4 x}~c\epsilon_{\m\n\r\s}F^{\m\n}F^{\r\s}~.
\label{deltaABBJ}
\end{equation}

The anomaly coefficient $r_{\rm p}$ does not get renormalizations \cite{adl-bard,PigSor}, it 
is identically zero if it vanishes at the one loop order, so it is sufficient to verify 
its vanishing at this order. However, that is the case, due to the fact that the potentially
dangerous axial current $j_5^\mu=\bar{\psi}\gamma_{5}\gamma^{\mu}\psi$ is 
only coupled to the external field $\b_\mu$ -- and not to any quantum field
of the theory -- which means that no gauge anomaly can be produced \cite{piguet-rouet,Ref11,Ref11'}. 
Hence it follows that the Slavnov-Taylor identity (\ref{quantum-slavnov}) is established at the quantum level.

Finally, in which concerns anomalies, the presence of a CPT violating interaction term by coupling 
an axial fermion current ($j_5^\mu=\bar{\psi}\gamma_{5}\gamma^{\mu}\psi$) with a constant vector 
field $b_\m$, does not induce neither a Lorentz anomaly nor a gauge anomaly -- independent of 
any regularization scheme. 

\subsection{The stability condition: in search for counterterms}
\hspace*{\parindent}
In order to verify if the action in the tree-approximation ($\widetilde{\Sigma}^{(s-1)}$) is stable 
under radiative corrections, we perturb it by an arbitrary integrated local functional (counterterm) 
$\widetilde\S^{c (s-1)}$, such that
\begin{equation}
\widehat\S^{(s-1)}=\widetilde\S^{(s-1)}+\ve \widetilde\S^{c (s-1)}~, \label{adef}
\end{equation}
where $\ve$ is an infinitesimal parameter. The functional $\widetilde\S^c\equiv\widetilde\S^{c (s-1)}|_{s=1}$ 
has the same quantum numbers as the action in the tree-approximation at $s=1$.

The deformed action $\widehat\S^{(s-1)}$ must still obey all the conditions presented above, henceforth, 
$\widetilde\S^{c (s-1)}$ is subjected to the following set of constraints:
\begin{eqnarray}
&&\cs_{\S^{(s-1)}}\widetilde\S^{c (s-1)}=0~, \label{stabcond}\\[3mm]
&&{\d\widetilde\S^{c (s-1)}\over{\d b}}={\d\widetilde\S^{c (s-1)}\over{\d{\ov c}}}=
{\d\widetilde\S^{c (s-1)}\over{\d c}}=0~, \label{cond}\\[3mm]
&&W_{\rm rigid} \widetilde\S^{c (s-1)}=0~, \label{crigidcond}\\[3mm]
&&{W}_{\rm L\a\b}\widetilde\S^{c (s-1)}=0~. \label{lorentzcond}
\end{eqnarray}

The most general invariant counterterm $\widetilde\S^{c (s-1)}$ -- the most general field polynomial -- with UV 
and IR dimensions bounded by $d\le4$ and $r\ge4$, with ghost number zero and fulfilling the conditions 
displayed in Eqs.(\ref{stabcond})-(\ref{lorentzcond}), reads:
\begin{align}
\widetilde\S^{c (s-1)}\bigg|_{r\ge4}^{d\le4}&=\int{d^4 x}~\biggl\{
\alpha_1 i\bar{\psi}\gamma^{\mu}(\partial_{\mu}+ieA_{\mu})\psi +  
\alpha_2 \bar{\psi}\psi + \alpha_3 F^{\mu\nu}F_{\mu\nu} +\nonumber\\[3mm]
&\quad+ \alpha_4 \left(\beta_{\mu}(x) + b_{\mu}\right)\bar{\psi}\gamma_{5}\gamma^{\mu}\psi \biggr\}~.
\label{finalcount}
\end{align}
The coefficients $\alpha_1,\ldots,\alpha_4$ are arbitrary, and they are fixed, order by order in perturbation 
theory, by the four normalization conditions:
\begin{eqnarray}
&&\G_{\bar\psi\psi}(\sl p)\bigg|_{{\ssl p}=m}=0~,~~
\frac{\partial}{\partial\sl p}\G_{\bar\psi\psi}(\sl p)\bigg|_{{\ssl p}=m}=1~,\nonumber \\[3mm]
&&\frac{\partial}{\partial p^2}\G_{A^TA^T}(p^2)\bigg|_{p^2=\kappa^2}=1~,~~
-\frac14 {\rm Tr}[\gamma^\mu\gamma^5 \G_{\b_\m\bar\psi\psi}(0,\sl p)]\bigg|_{{\ssl p}=m}=1~.
\label{normcond}
\end{eqnarray}

It shall be stressed here that, a Chern-Simons-like term of the type
\begin{equation}
\S_{\rm CS}=\int{d^4 x}~\alpha_5\biggl\{\epsilon_{\mu\nu\alpha\beta}\beta^{\mu}(x)A^{\nu}\partial^{\alpha}A^{\beta}\bigg|_4^4
+ \epsilon_{\mu\nu\alpha\beta}b^\mu A^{\nu}\partial^{\alpha}A^{\beta}\bigg|_3^3\biggr\}~, 
\label{CS}
\end{equation}
in spite of fulfils the conditions (\ref{cond})-(\ref{lorentzcond}), its first term breaks gauge invariance by 
violating the Slavnov-Taylor identity (\ref{stabcond}), whereas its second term violates IR dimension 
constraint ($r_{\S_{\rm CS}}\ge4$), it has IR dimension equal to three. Therefore, the Chern-Simons-like term 
$\S_{\rm CS}$ ($\ref{CS}$) can never be generated by radiative corrections if the renormalization procedure 
is performed correctly. First, by taking care of the IR divergences -- for instance, through the 
Lowenstein-Zimmermann method~\cite{Low} -- that show up, thanks to the presence of a photon, which is massless. 
Second, by properly treating and controlling the Lorentz symmetry breaking through the Symanzik method~\cite{Sym1,Sym2}. 
Anyway, even though the external field $\beta_\mu(x)$ was not introduced in order to control the Lorentz breaking, 
the Chern-Simons-like term -- which is a soft Lorentz breaking (UV dimension less than four) -- would not be 
radiatively generated as explained above, nevertheless, any gauge invariant hard Lorentz breaking (UV dimension 
equal to four) could be induced by radiative corrections. In short, a CPT-odd and Lorentz-violating Chern-Simons-like 
term, {\it independent of any regularization scheme}, is definitely {\it not} radiatively induced by  coupling an axial fermion 
current ($j_5^\mu=\bar{\psi}\gamma_{5}\gamma^{\mu}\psi$) with a constant vector field $b_\m$.

\section{\bf Conclusions}
\label{Sec5}
\hspace*{\parindent}
We revisit the issue intensively studied in recent years on the generation of terms by radiative corrections in models with 
broken Lorentz symmetry. We have exemplified all of this in two examples. In  the first example, we have discussed the 
case of the hard breaking for a very simple model involving scalar fields and tried to give some insight on the way 
Symanzik's method of introducing controlling exterior fields is working. We have explicitly shown the difference of 
theory's behaviour, the external field being present or being not present. An interesting collateral result is the presence 
in this model of an infinite number of counterterms -- all compatible with power-counting renormalizability -- which could 
jeopardize renormalizability in the sense of needing an infinite number of normalization conditions to fix them, which physically 
corresponds to an infinite set of measurements. But we have argued that, because the Lorentz violation, if physically present, 
must be very small, and thus only a few of these counterterms is practically relevant. In  the second example, we reassess the 
discussion on the radiative generation of a Chern-Simons-like term induced from quantum corrections in the extended QED.
We show, to all orders in perturbation theory, that a CPT-odd and Lorentz violating Chern-Simons-like term, 
{\it definitively}, is not radiatively induced by the axial coupling of the fermions with the constant vector $b_\mu$.
The proof of this fact is based on general theorems of perturbative quantum field theory, where the 
Lowenstein-Zimmermann subtraction scheme in the framework of Bogoliubov-Parasiuk-Hepp-Zimmermann-Lowenstein 
(BPHZL) renormalization method is adopted.

It is true that we need new ideas to go beyond the Standard Model. An idea so is the Lorentz symmetry breaking. 
If it is present in our universe has been the subject of much discussion. So far, no trace was found. Experience is 
the final judgment of a theory; therefore to be checked experimentally, the Lorentz symmetry breaking 
remains a theoretical construction, regardless of how seductive the idea can be. However, even as a theoretical construction, 
the idea of the Lorentz symmetry breaking should be well grounded. But it seems that this has not happened in the recent 
literature on the subject. In particular, in this article we analyze the issue intensively studied in recent years on the generation 
of terms by radiative corrections in models with broken Lorentz symmetry. Unfortunately, several recent works, dealing the 
subject, do not consider very carefully how the Lorentz symmetry is broken, not taking into account 
the requirements that  Symanzik-Becchi-Rouet-Stora have shown to be necessary. The young researchers who study a QFT 
with broken Lorentz symmetry should read the pioneer articles of Symanzik-Becchi-Rouet-Stora and ``devour them.'' 
Exactly what we do over the years! And that is why we hope to have conveyed the impression that the reconsideration of 
the fundamental works on renormalization of quantum field models developed mainly in the 1970's, especially the papers of 
Symanzik-Becchi-Rouet-Stora on renormalizable models with broken symmetry, provide us with a theoretical tool 
susceptible to avoid some ``bad'' conclusions associated with models with broken Lorentz symmetry. It is important to 
emphasize that, the main characteristics of this method is the {\it control} of the breaking and, in particular, its power-counting 
properties, converting the initial action containing terms that violate the Lorentz symmetry into one which is invariant under 
the original transformation adding external fields (the Symanzik sources). Without this control, the study of the stability 
(here meant additive renormalization) tells us that {\it any} term that breaks the Lorentz symmetry, compatible with the 
power-counting, must {\it necessarily} be present in the starting lagrangian. On the other hand, if we include in the initial 
lagrangian all terms that break the symmetry Lorentz, compatible with the locality and power-counting, no breaking control 
is required (see Ref.~\cite{Nos2}). Therefore, paraphrasing Symanzik, {\it whether you like it or not, you have to include in the 
initial lagrangian all terms that violate the Lorentz symmetry consistent with locality and power-counting, unless otherwise 
constrained by a break control!}

\section*{Acknowledgments} 
\hspace*{\parindent}
This work was partially funded by the Funda\c{c}\~ao de Amparo \`a Pesquisa do Estado de Minas Gerais -- 
FAPEMIG, Brazil (O.P.) and the Conselho Nacional de Desenvolvimento Cient\'{\i}fico e Tecnol\'{o}gico -- CNPq, Brazil (O.P.).



\end{document}